\documentclass[aps,pre,twocolumn,superscriptaddress]{revtex4-2}
\usepackage[usenames,dvipsnames]{xcolor}
\usepackage{amsmath,amssymb,graphicx,bm}

\begin{document}

\title{Emergence of Stigmergic Transport in Granular Environments}

\author{F.Wéry} 
\affiliation{GRASP, Physics Department, University of Li\`ege, B-4000 Li\`ege, Belgium}
\author{F.N. Pinan Basualdo} 
\affiliation{Department of Mechanical Engineering, Katholieke Universiteit Leuven, 3001 Leuven, Belgium}
\author{B. Gorissen} 
\affiliation{Department of Mechanical Engineering, Katholieke Universiteit Leuven, 3001 Leuven, Belgium}
\author{N.Vandewalle} 
\affiliation{GRASP, Physics Department, University of Li\`ege, B-4000 Li\`ege, Belgium}

\date{\today}

\begin{abstract}
We show that stigmergic path formation emerges in a deformable environment through the interplay between environmental memory and geometrical crowding. Using experiments with robotic random walkers together with a minimal stochastic model, we demonstrate the onset of persistent self-reinforced transport pathways above a critical packing fraction $\phi_c$, where environmental memory enhances walker mobility. As the jamming transition $\phi_J$ is approached, increasing crowding progressively suppresses this transport enhancement despite the persistence of environmental memory. The resulting non-monotonic behavior reveals an optimal transport regime well below jamming. More generally, our work establishes how active agents can collectively build transport networks through purely mechanical interactions with a deformable substrate.
\end{abstract}

\maketitle

Collective path formation is one of the most remarkable examples of environment-mediated self-organization. In biological systems, this phenomenon is known as \emph{stigmergy} \cite{stigm,stigm2}, whereby agents coordinate their activity indirectly through persistent traces left in a shared environment. Originally introduced to describe nest construction by social insects such as ants \cite{ants} and termites \cite{termites}, and later extended to human trail formation \cite{helbing,helbing2} and robotic systems\cite{termes}, stigmergy has become a unifying framework for a broad class of systems in which the environment mediates interactions between moving agents. Rather than acting as a passive background, the substrate simultaneously stores information and modifies subsequent transport.

Recent studies have highlighted the fundamental role of environment-mediated transport in crowded and deformable media. Similar questions also arise in the physics of active crowds, where self-propelled robots evacuating through constrictions reproduce the collective dynamics of pedestrian flows and granular bottlenecks \cite{ZuriguelPRL2026}. While transport in disordered systems is traditionally understood from a percolation perspective \cite{StaufferAharony}, allowing moving agents to rearrange obstacles qualitatively changes the transport properties of the medium \cite{sokobanwalker1,sokobanwalker2,pushywalker}. Similar behaviors have recently been observed in active granular systems, where robots or active filaments continuously reshape their surroundings by collecting particles, forming clusters, or carving cavities \cite{altshuler,brun2024,brun2025,boudet2021,deblais2026}. These studies, together with recent experiments on robotic and human evacuations through bottlenecks \cite{ZuriguelPRL2026}, demonstrate that agents continuously remodel their environment. However, they primarily address how agents modify the substrate, leaving largely unexplored the reciprocal process by which the remodeled substrate subsequently enhances transport.

How can environmental remodeling become self-reinforcing and eventually generate persistent transport pathways? This question has recently attracted renewed attention in active matter, where trail-mediated transport has been reported for wave-memory systems \cite{hubert}, active colloids \cite{dias}, and robotic walkers \cite{altshuler}. In all these systems, the trajectory of a moving agent modifies the environment, which in turn biases future trajectories, generating a feedback loop between motion and substrate memory.

Most existing works implicitly assume that stronger environmental memory necessarily promotes more efficient path formation. While this picture is appropriate for many systems, it remains incomplete for granular-like or crowded environments. In such media, a footprint is not simply written and read. It may also be screened by local rearrangements, erased by subsequent passages, or become inaccessible because of steric constraints. Path formation therefore results from the competition between two antagonistic mechanisms: environmental memory reinforces previously visited trajectories, whereas crowding progressively suppresses mobility and may ultimately frustrate the very mechanism that memory is expected to promote.

\begin{figure}[h]
    \centering
    \includegraphics[width=0.95\linewidth]{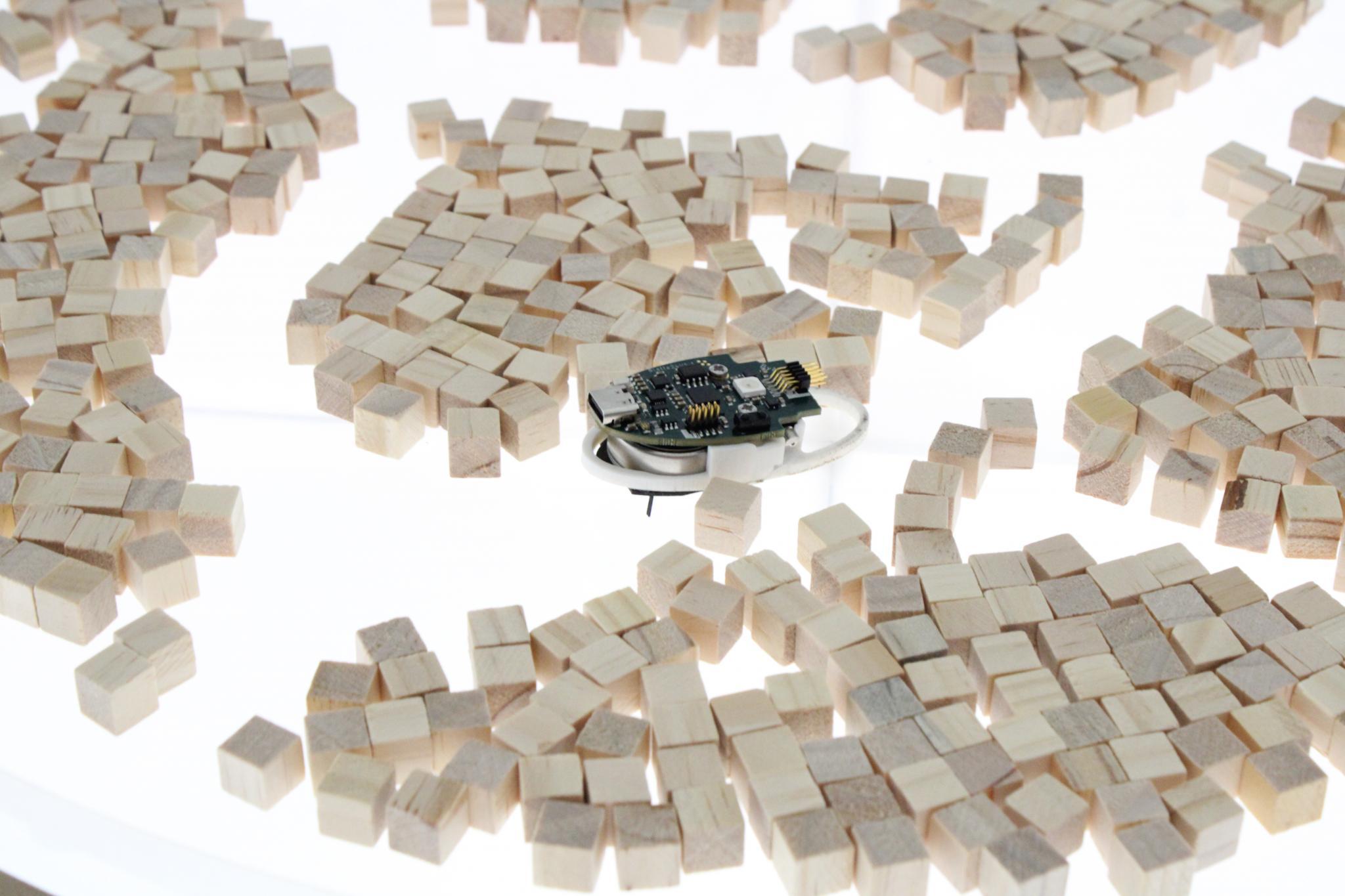}
    \caption{A bristlebot (GRASPion, \cite{rsi}) is forming traces by pushing cubic wood blocks in an arena. Light comes from below. }
    \label{fig:intro}
\end{figure}

Packing fraction therefore emerges as the natural control parameter governing this competition. At low packing fraction, the substrate is too weakly structured to retain persistent traces, and the walker performs an almost random exploration. At high packing fraction, traces become increasingly persistent, but mobility is progressively reduced by steric constraints, eventually leading to jamming. These opposing tendencies suggest an intermediate regime in which environmental memory is sufficiently long-lived, yet walkers remain mobile enough to repeatedly exploit the pathways they create. In this work, we demonstrate experimentally and numerically that this balance governs the onset of stigmergic transport in a deformable granular environment.

We placed thousands of small wooden cubes of volume $1\,{\rm cm}^3$ on a square arena of size $100\times100\,{\rm cm}^2$. The cubes were initially dispersed uniformly by hand to reach a prescribed average packing fraction $\phi$, varied from $0.00$ to $0.65$ in steps of $0.05$. An elliptical bristlebot (GRASPion~\cite{rsi}) of size $6\times3\,{\rm cm}^2$ was then introduced and programmed to perform a random walk with a persistence time of $0.8\,{\rm s}$, corresponding to a persistence length of about $10\,{\rm cm}$. To avoid boundary-following effects, the robot was stopped and reoriented toward the interior whenever it reached the arena boundary. An overhead camera simultaneously recorded the robot trajectory and the evolution of the granular assembly. For image analysis, the arena was partitioned into a $30\times30$ grid of square cells, whose size approximately matched the minor axis of the robot. The local packing fraction $\phi_i$ was measured in each cell, with $\langle\phi_i\rangle=\phi$. For each value of $\phi$, at least three independent experiments were performed.

\begin{figure}
    \centering
    \includegraphics[width=1\linewidth]{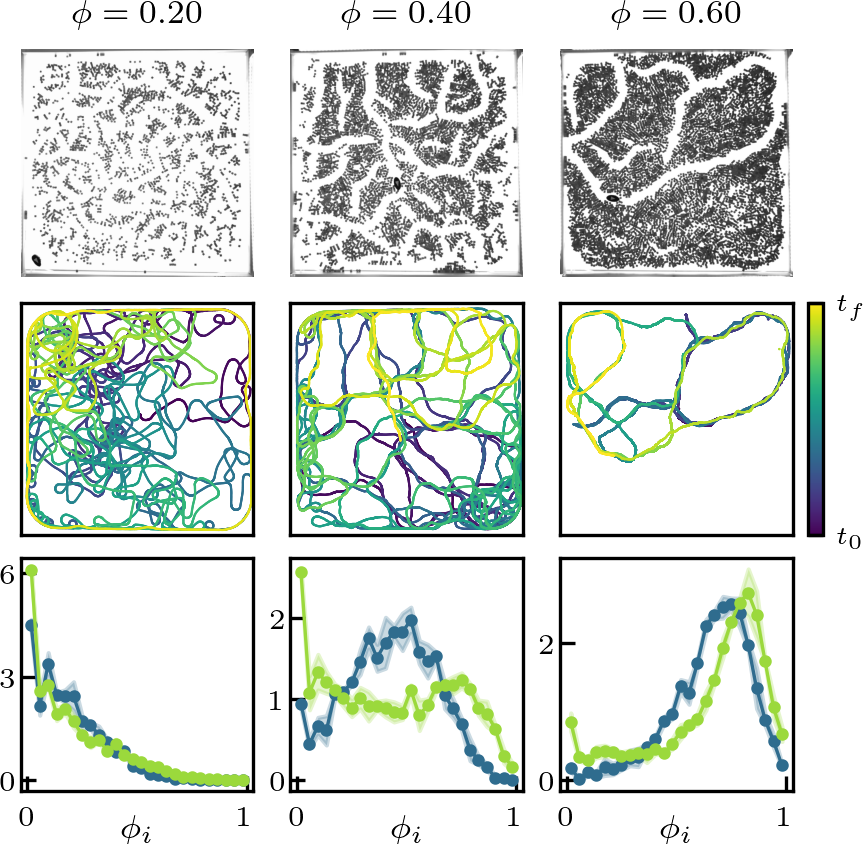}
    \caption{Main experimental results for three packing fractions, shown from left to right: $\phi=0.2$, $\phi=0.4$, and $\phi=0.6$. (top row) Images of the arena, emphasizing particle organization and trail formation at intermediate/high $\phi$. (second row) Trajectories of the robot, with time indicated by the color scale. Duration $t_f-t_0$ is 12 min (low $\phi$) and 20 min otherwise. Traces reveal multiple passages at intermediate/high $\phi$. (bottom row) Probability Density Functions (PDF) of the local packing fraction $\phi_i$ when a $30 \times 30$ grid is considered on images. Blue curves correspond to the initial state $t_0$, while green curves correspond to the final state $t_f$.}
    \label{fig:experimental}
\end{figure}

The top row of Fig.~\ref{fig:experimental} shows representative snapshots of the system after $12$ to $20$ minutes of motion, for three different packing fractions. At low $\phi$, the trajectories of the walker remain diffuse, and no robust trail emerges. The observed pattern is mainly governed by the random motion of the bot, as illustrated in the second row of the figure. At intermediate values of $\phi$, paths are progressively reinforced by repeated visits. The bot is then observed to preferentially follow some of the previously formed trails. At high $\phi$, the medium becomes increasingly constrained: the motion is hindered, the footprints are spatially frustrated, and the resulting trajectories become less efficient at creating new trails. However, once a trail has been formed, the bot tends to follow the same traces repeatedly, as seen in the second row of Fig.~\ref{fig:experimental}.
This behavior contrasts with recent studies of active carving in granular media \cite{pushywalker, brun2025}, where displaced material accumulates ahead of the walker or around a cleared region, eventually producing compact cavities and confinement. Here, by contrast, obstacles are predominantly displaced sideways from the walker trajectory. Repeated passages therefore reinforce elongated depleted corridors rather than isotropically expanding cavities. Once formed, these corridors offer lower geometrical resistance and are preferentially revisited, closing the feedback loop between environmental remodeling and walker motion.

The bottom row of Fig.~\ref{fig:experimental} shows the Probability Density Function (PDF) of the local packing fractions $\phi_i$, measured from the images. The blue curve corresponds to the initial distribution, which is approximately Gaussian-like around the global packing fraction $\phi$, although with broad tails. Trail formation leaves nearly empty sites behind the bot and produces regions of higher packing fraction along the sides of the trails. This is visible in the green curve, corresponding to the final state of the experiment, where a peak appears close to $\phi_i=0$ (path formation) and at larger values of $\phi_i> \phi$ as denser regions appear.

\begin{figure}[t]
    \centering
    \includegraphics[width=1\linewidth]{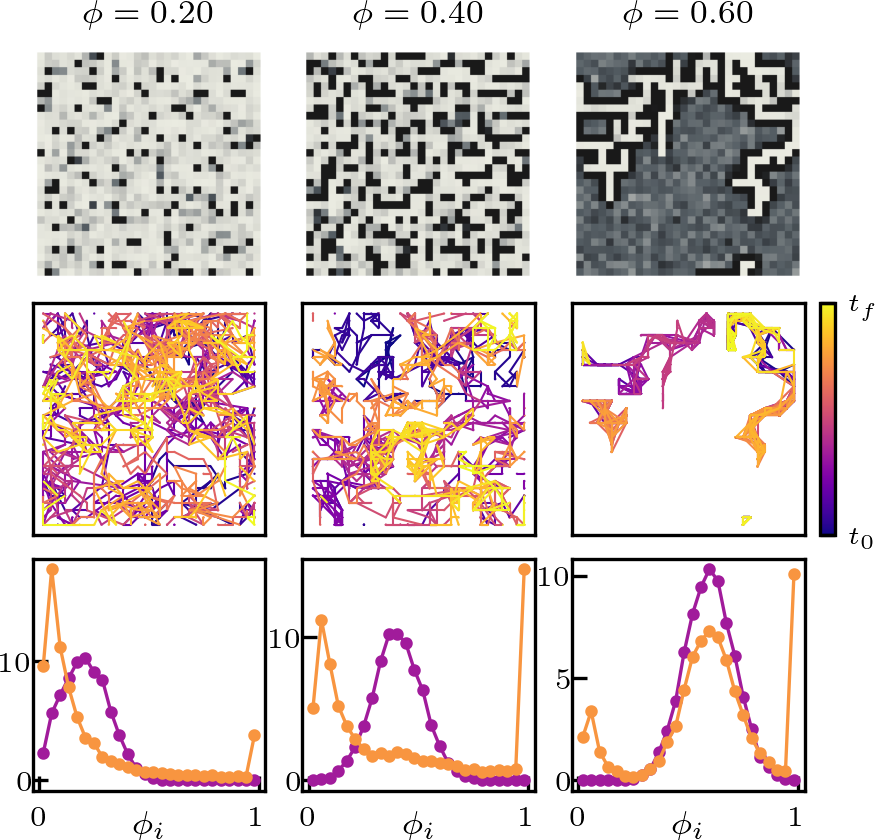}
    \caption{Main numerical results for three packing fractions, shown from left to right: $\phi=0.2$, $\phi=0.4$, and $\phi=0.6$. First row: Images of the system at the three packing fractions. Second row: Trajectories of the robot, with time indicated by a color scale. Third row: Probability Density Functions (PDF) of the local packing fraction $\phi_i$. Violet curves correspond to the initial state, while orange curves correspond to the final state.}
    \label{fig:numerical}
\end{figure}

To rationalize our observations, we introduce a minimal coarse-grained model of a random walker moving in a dynamically evolving environment. Our model differs from the Pushy Random Walk introduced in Ref.~\cite{pushywalker} in an essential aspect. In the latter, obstacles are displaced primarily along the direction of motion, so that the walker progressively pushes its environment ahead. Here, material displaced by the walker is redistributed among neighboring cells, including transverse directions. This lateral redistribution allows depleted regions to persist behind successive passages and therefore provides a minimal mechanism for trail formation. As in the image analysis, the arena is partitioned into a $L\times L$ lattice of cells. Each lattice site $i$ does not represent an individual cube but a coarse-grained region characterized by its local packing fraction $\phi_i$, with a prescribed average packing fraction $\phi=\langle\phi_i\rangle$. At time $t$, the walker occupies site $i$. At the next time step, it selects one of the neighbouring sites $j$ at random and moves to it with probability
\begin{equation}
P_{ij}=1-\phi_j,
\label{eq:proba}
\end{equation}
otherwise remaining at site $i$. The walker therefore preferentially explores locally dilute regions. After each successful move, the newly occupied site is partially depleted, and the displaced material is redistributed isotropically among its neighboring cells, while enforcing $\phi_i\le1$. Periodic boundary conditions are used throughout the simulations.

As shown in Figure \ref{fig:numerical}, numerical simulations reproduce the three main regimes observed experimentally as $\phi$ increases: random exploration at low packing fraction close to a classical Random Walk (RW), trail formation at intermediate $\phi$, and narrow path formation visited multiple times at large packing fraction. No directional persistence is imposed, showing that trail formation can emerge solely from the coupling between walker motion and substrate redistribution. Similar to experiments, Figure \ref{fig:numerical} displays the PDF of $\phi_i$. In the path-forming regime, the distribution becomes strongly heterogeneous, developing populations of depleted and densely packed sites that reflect the redistribution induced by repeated walker passages. 

\begin{figure}
    \centering
    \includegraphics[width=0.95\linewidth]{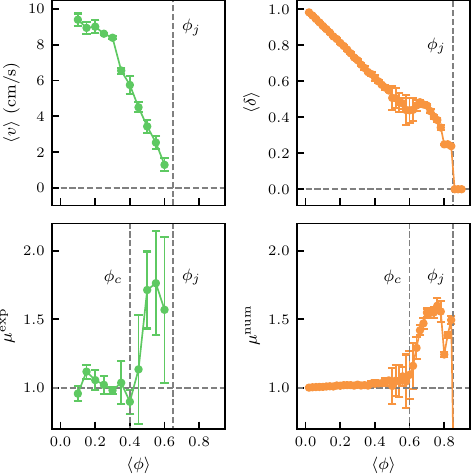}
    \caption{Speed and mobility of bot/walker in both experiments (left column) and simulations (right column). Critical points are denoted by vertical dashed lines. Speed vanishes to zero at $\phi^{\rm exp}_J \approx 0.65$ in experiments. Mean displacement per iteration $\delta$ reaches zero at $\phi^{\rm num}_J \approx 0.85$ in simulations. Mobility $\mu^{\rm exp}$ of the bot as a function of $\phi$, showing excess mobility above $\phi_c=0.40$. Numerical mobility $\mu^{\rm num}$ also shows excess above $\phi_c \approx 0.60$.}
    \label{fig:4}
\end{figure}

From the experimental and numerical observations presented above, it appears that stigmergic behaviour emerges only above a finite packing fraction. We now quantify this onset using two independent observables. We first consider the walker dynamics by measuring its average velocity throughout the experiment. As shown in Fig.~\ref{fig:4}, the average velocity $\langle v\rangle$ continuously decreases with increasing packing fraction and eventually vanishes at $\phi_J^{\rm exp}\approx0.65$, identifying the jamming transition. At this point, large compact aggregates prevent the bot from rearranging the surrounding blocks, and motion ceases. In the numerical model, complete jamming occurs at the larger value $\phi_J^{\rm num}\approx0.85$, reflecting the simplified interaction rules. Despite this quantitative shift, both systems exhibit the same qualitative behaviour.

To isolate the effect of trail formation from the trivial slowdown caused by crowding, we compare the velocity during the initial stage of the experiment, before trails have developed, $\langle v_0\rangle$, with the velocity measured once a stationary state is reached. We define the normalized mobility
\begin{equation}
\mu^{\rm exp}=\frac{\langle v\rangle}{\langle v_0\rangle}.
\end{equation}
For a purely random walk, $\mu\simeq1$. Figure~\ref{fig:4} reveals a clear increase of $\mu$ above $\phi_c^{\rm exp}\approx0.40$, precisely where persistent trails first appear. Once established, these trails increase the walker mobility by nearly $50\%$. Rather than simply storing information, the environment becomes progressively reorganized into preferential transport pathways that facilitate subsequent motion.

The same mechanism is reproduced by the numerical model. There, the average displacement per iteration, $\langle\delta\rangle$, plays the role of the walker velocity. As expected from Eq.~(\ref{eq:proba}), $\langle\delta\rangle$ decreases with increasing packing fraction and vanishes at $\phi_J^{\rm num}\approx0.85$. Removing this purely geometrical effect by normalizing with the available free space,
\begin{equation}
\mu^{\rm num}=\frac{\langle\delta\rangle}{1-\phi},
\end{equation}
reveals the same mobility enhancement around $\phi_c^{\rm num}\approx0.60$. The numerical model therefore reproduces the experimentally observed self-induced acceleration associated with trail formation.

\begin{figure}
    \centering
    \includegraphics[width=0.85\linewidth]{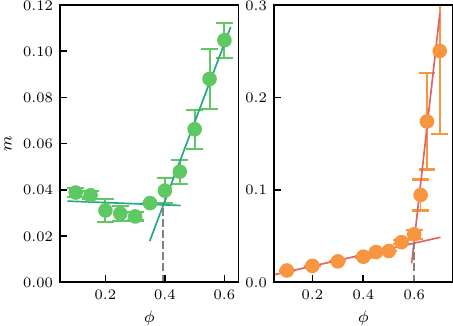}
    \caption{Memory order parameter $m$, defined in Eq. (\ref{eq:memory}), as a function of the packing fraction $\phi$. The sharp increase at $\phi_c$, denoted by a dashed vertical line, identifies the onset of stigmergic path formation in both experiments and simulations.}
    \label{fig:5}
\end{figure}

Although the mobility already provides a clear signature of the emergence of persistent transport pathways, it remains sensitive to velocity fluctuations. We therefore introduce a second observable that directly quantifies the memory encoded in the environment. Let $p_i$ denote the probability that the walker occupies grid cell $i$. For a purely random walk, one expects a uniform occupation probability, $p_i=1/L^2=p_0$, for every cell. We therefore define the memory order parameter
\begin{equation}
m=\sqrt{\sum_i\left(p_i-p_0\right)^2},
\label{eq:memory}
\end{equation}
which vanishes for perfectly homogeneous exploration and increases as repeated passages become concentrated along preferential paths. Because the experimental trajectories have a finite duration, the measured probabilities $p_i$ exhibit statistical fluctuations even in the random-walk regime. Consequently, $m$ remains finite at low packing fractions, with a baseline value of approximately $0.04$ for $L=30$, as observed in Fig.~\ref{fig:5}. Above $\phi_c^{\rm exp}\approx0.40$, however, $m$ increases sharply, marking the onset of stigmergic transport. In the numerical simulations, much longer trajectories reduce these finite-sampling fluctuations, yielding values of $m$ closer to zero at low packing fractions and a clear increase above $\phi_c^{\rm num}\approx0.60$. These results demonstrate that stigmergic transport requires a finite degree of crowding, whereas excessive crowding ultimately suppresses transport through jamming.

Taken together, these results reveal that stigmergic path formation is governed by the competition between environmental memory and geometrical constraints. The environment is not merely a passive information storage medium. Instead, it evolves into a self-organized transport network that simultaneously records previous trajectories and enhances future motion. Maximum mobility enhancement is achieved well before the jamming transition, where memory is sufficiently persistent while mobility remains possible.

\vskip 2cm

We have investigated stigmergic path formation in a granular-like environment using experiments together with a minimal stochastic model. Both approaches reveal a non-monotonic dependence on packing fraction. At low density, environmental modifications are too weak to sustain persistent trails. At high density, geometrical constraints suppress mobility and ultimately lead to the jamming transition. Between these two limits lies an optimal regime where environmental memory and walker mobility combine to generate robust self-organized transport pathways.

More generally, our work suggests that transport in deformable environments cannot be understood independently of the medium that agents continuously remodel. Rather than simply adapting to their surroundings, active agents collectively construct the transport network that subsequently guides their own motion. This mechanism should apply broadly whenever locomotion and environmental remodeling are intrinsically coupled, from robotic swarms and pedestrian traffic to active matter and biological collectives. Future work could investigate how the strength of environmental remodeling controls the onset of stigmergy, extend the coarse-grained model toward more realistic substrate dynamics, and explore collective transport mediated by self-generated trails, potentially leading to clustering phenomena reminiscent of those observed in active colloidal systems~\cite{dias}.

\section*{Acknowledgements}

This work is financially supported by the University of Liège through the CESAM Research Unit. N.V. thanks the Fondation Francqui for its support. F.N.P.B. and B.G. thank the European Commission for its support under the Horizon Europe program under Grant \#101076036 (ILUMIS).

F.W. and F.N.P.B. contributed equally to this work.


\begin{thebibliography}{99}

\bibitem{stigm} P.-P. Grassé, 
{\it La théorie de la stigmergie},
Insectes Sociaux {\bf 6}, 41 (1959)

\bibitem{stigm2} G.Theraulaz and E.Bonabeau, 
{\it A Brief History of Stigmergy},
Artif Life {\bf 5}, 97–116 (1999)

\bibitem{ants} A.Khuong, J.Gautrais, A.Perna, C.Sbaï, M.Combe, P.Kuntz, Ch.Jost, and G.Theraulaz,
{\it Stigmergic construction and topochemical information shape ant nest architecture},
PNAS {\bf 113}, 1303–1308 (2016)

\bibitem{termites} S.A.Ocko, A.Heyde, and L.Mahadevan
{\it Morphogenesis of termite mounds},
PNAS {\bf 116}, 3379–3384 (2019) 

\bibitem{helbing} D.Helbing, J.Keltsch, P.Molnár, 
{\it Modelling the evolution of human trail systems},
Nature {\bf 388}, 47–50 (1997)

\bibitem{helbing2} D.Helbing, F.Schweitzer, J.Keltsch, P.Molnár, 
{\it Active walker model for the formation of human and animal trail systems}, 
Phys. Rev. E {\bf 56}, 2527 (1997)

\bibitem{termes} J. Werfel, K. Petersen, and R. Nagpal
{\it Designing collective behavior in a termite-inspired robot construction team},
Science {\bf 343}, 754--758 (2014) 

\bibitem{ZuriguelPRL2026} M.Grasa, L.Alonso-Llanes, A.Garcimartín, and I.Zuriguel, 
{\it Flow through Bottlenecks: Stronger Is Slower},
Phys. Rev. Lett. {\bf 137}, 058301 (2026)

\bibitem{StaufferAharony}
D. Stauffer and A. Aharony,
{\it Introduction to Percolation Theory},
Taylor \& Francis, London (1992)

\bibitem{sokobanwalker1} 
O. Lauber Bonomo and S. Reuveni,  
{\it Loss of percolation transition in the presence of simple tracer-media interactions},  
Phys. Rev. Res. {\bf 5}, L042015 (2023)

\bibitem{sokobanwalker2} 
P. Singh, D. A. Kessler, and E. Barkai,  
{\it Sokoban random walk: From environment reshaping to trapping crossover},  
Phys. Rev. Res. {\bf 8}, L012023 (2026)

\bibitem{pushywalker} 
O.Lauber Bonomo, I.Shitrit, S.Reuveni, and S.Redner,  
{\it Pushy Random Walk: A Minimal Model for Transport in Deformable Media},  
Phys. Rev. Lett. {\bf 137}, 037101 (2026)

\bibitem{altshuler} 
A.Altshuler, O.Lauber Bonomo, N.Gorohovsky, S.Marchini, E.Rosen, O.Tal-Friedman, S.Reuveni, and Y.Roichman,
{\it Environmental memory facilitates search with home returns},
Phys. Rev. Res. {\bf 6}, 023255 (2024)

\bibitem{brun2024}
Y.Xi , T.Marzin, R.B.Huang, and P.-T.Brun, 
{\it Emergent behaviors of buckling-driven elasto-active structures}, 
PNAS {\bf 121}, e2410654121 (2024)

\bibitem{brun2025}
Y. Xi, T. Marzin, and P.-T. Brun,
{\it Building granular structures with elasto-active systems},
arXiv:2511.01378 (2025)

\bibitem{boudet2021}
J.-F. Boudet, J. Lintuvuori, C. Lacouture, T. Barois, A. Deblais,
K. Xie, S. Cassagnere, B. Tregon, D. B. Brückner, J.-C. Baret, {\it et al.},
{\it From collections of independent, mindless robots to flexible, mobile, and directional superstructures},
Sci. Robot. {\bf 6}, eabd0272 (2021)

\bibitem{deblais2026}
R. Sinaasappel, K. R. Prathyusha, H. Tuazon, E. Mirzahossein,
P. Illien, S. Bhamla, and A. Deblais,
{\it Particle sweeping and collection by active and living filaments},
Phys. Rev. X {\bf 16}, 011003 (2026)

\bibitem{hubert} M.Hubert, S.Perrard, N.Vandewalle and M.Labousse, 
{\it Overload wave-memory induces amnesia of a self-propelled particle}, 
Nat. Comm. {\bf 13}, 4357 (2022)

\bibitem{dias} Environmental memory boosts group formation of clueless individuals, 
C.S.Dias, M.Trivedi, G.Volpe, N.A.M.Araújo, G.Volpe,
Nat. Comm. {\bf 14}, 7324 (2023)

\bibitem{rsi} F.Novkoski, M.Mélard, M.Delens, F.Wéry, M.Noirhomme, J.Pande, A.Maier, A-S.Smith, N.Vandewalle, 
{\it GRASPion: An open-source, programmable brainbot for active matter research}, 
Rev. Sci. Instrum. {\bf 97}, 014704 (2026)


\end{thebibliography}
\end{document}